\documentclass[aps,prb,twocolumn,showpacs,eqsecnum,superscriptaddress,floatfix,10pt]{revtex4-1}
\usepackage{bm,amsmath,amssymb,gensymb,amsfonts,stmaryrd,wasysym,graphicx,multirow,color,textcomp,braket}
\usepackage{url}
\usepackage{graphicx}
\usepackage[caption=false,singlelinecheck=false]{subfig}
\usepackage[colorlinks,linkcolor=blue,citecolor=cyan,filecolor=magenta,urlcolor=cyan]{hyperref}

\newcommand{\nn}{\nonumber}

\begin{document}

\title{Partial Kondo Screening and Anomalous Thermal Conductance}
\author{Hee Seung Kim}
\author{Hyeok-Jun Yang}
\author{SungBin Lee}
\affiliation{Department of Physics, Korea Advanced Institute of Science and Technology, Daejeon, 34141, Republic of Korea
}
\date{\today}

\begin{abstract}
The frustrated magnetism on the Kondo lattice system motivates intriguing Kondo-breakdown beyond the traditional Doniach scenario. Among them, the fractionalized Fermi liquid (FL*) has drawn a particular interest by virtue of its fractionalized nature. Here, we study the phase diagram of $J_{1}$-$J_{2}$ Kondo-Heisenberg model on a honeycomb lattice at a quarter filling. Employing the slave-fermion mean-field theory with $d \pm id$ spin liquid ansatz and exact diagonalization, we discuss the emergence of partial Kondo screening in the frustrated regime with comparable $J_{1}$ and $J_{2}$, and the fractionalized superconductor (SC*) which is superconductor analogy of the FL*. Due to the larger number of local spin moments than itinerant electrons, the magnetic fluctuation is still significant even in the strong-coupling limit, which influences the thermodynamic and transport properties qualitatively. In particular, we estimate the thermal conductance to probe the low-energy excitation and show the anomalous behaviour in the SC* phase contrast to the conventional superconductors.
\end{abstract}

\maketitle

\section{Introduction}
Kondo lattice model (KLM) is prototypical for designing the heavy fermion compounds and transition-metal oxides\cite{RevModPhys.56.755,hewson_1993,RevModPhys.79.1015}. After the discovery of resistivity minimum\cite{DEHAAS19341115,10.1143/PTP.32.37,RevModPhys.40.380}, a wealth of correlated phenomena such as the effective mass enhancement\cite{PhysRevLett.35.1779}, quantum criticality\cite{Maple1994,Coleman_2001,PhysRevLett.89.056402,Paschen2004,RevModPhys.78.743,Gegenwart2008}, and unconventional superconductivity\cite{Mathur1998,RevModPhys.81.1551,Steglich_2012,PhysRevLett.110.146406} have been reported. A great deal of theoretical interest is devoted to understand those novel characters as a consequence of two competing tendencies, Ruderman-Kittel-Kasuya-Yosida (RKKY) interaction and the resonant hybridization on the KLM\cite{RevModPhys.48.219,DONIACH1977231}. The underlying paradigm is that the dense arrays of magnetic impurities immersed in the metal undergo several instabilities, especially the spin density waves and Kondo singlet formation as the ratio of the competing energy scales varies below the Kondo temperature.

Recently, it has been spotlighted that the geometrical frustration is also decisive to the global phase diagram in Kondo physics\cite{PhysRevB.66.045111,PhysRevLett.103.066405,Coleman2010,https://doi.org/10.1002/pssb.201390006}. In the frustrated magnet, the low-temperature paramagnet might develop a highly entangled ground state, quantum spin liquid (QSL) which turns out to be robust against small perturbations\cite{Savary_2016_1}. Then, the Kondo coupling to the conduction electrons might lead to interesting physics, fractionalized Fermi liquid (FL*)\cite{PhysRevLett.90.216403,PhysRevB.69.035111}. Compared to conventional Fermi liquid (FL), the robustness against the Kondo screening is supported by the topological order, accompanied with the Luttinger theorem violation. An alternative possibility for the small Fermi surface is the partial Kondo screening (PKS) separated from the strong coupling limit to some extent\cite{PhysRevLett.105.036403,PhysRevLett.120.107201}. In the presence of the intermediate Kondo exchange, only a portion of local moments is hybridized leaving the magnetic degrees of freedom. In the thermodynamic limit, it has been speculated that the portion of singlets tends to be regularly arrange on a specific sublattice.

In this way, the frustrated magnetism involves both FL* and PKS, but their collaboration on KLM leaves an open question. Starting from FL* instead of magnetic order, it is tempting to consider the coexisting characters of FL* and PKS as the Kondo coupling is turned on. If the numbers of Bloch electrons and localized $S = 1/2$ spins are stringently equal, the ground state inevitably flows to the heavy Fermi liquid in the strong coupling limit. Then the local spin moments altogether engages in the hybridization channel. However, if there is an excess of magnetic degrees of freedom, the strongly-coupled KLM does not necessarily implies the Kondo limit. Along with the electronic quasi-particles, the remaining magnetic spins still strongly fluctuate to bring about the charge-neutral spinons. As a consequence, the transport and thermodynamic behaviours are expected to exhibit both the fractionalization and hybridization characters.

In this paper, we consider $J_{1}$-$J_{2}$ Kondo-Heisenberg model on the honeycomb lattice with quarter-filled conduction electrons. Employing the ground state ansatz of frustrated Heisenberg model to be a $d \pm id$ spin liquid, we turn on the onsite Kondo exchange to propose the existence of a novel fractionalized phase beyond FL*. As the Kondo exchange increases, we show that only one sublattice is spontaneously hybridized forming a Kondo resonance, while the other sublattice still retains the fractionalized excitation. Especially for sufficiently frustrated regime where $J_{1}$ and $J_{2}$ are comparable, the proximate superconductivity is induced preserving the chirality of the superconducting order on the itinerant side. As a consequence of existing gapless fractionalized excitation, we expect the power-law longitudinal thermal conductance behavior at the low temperature. This result is contrary to the $s$-wave or chiral $d$-wave superconductors with gapped spectrum exhibiting exponentially decaying behaviors. Finally, we also check that the partial Kondo screening occurs for exact diagonalization on the $18$-sites honeycomb lattice.

\section{$J_{1}$-$J_{2}$ Kondo-Heisenberg model on the honeycomb lattice}
\begin{figure}
	\subfloat[]{\includegraphics[width=0.23\textwidth]{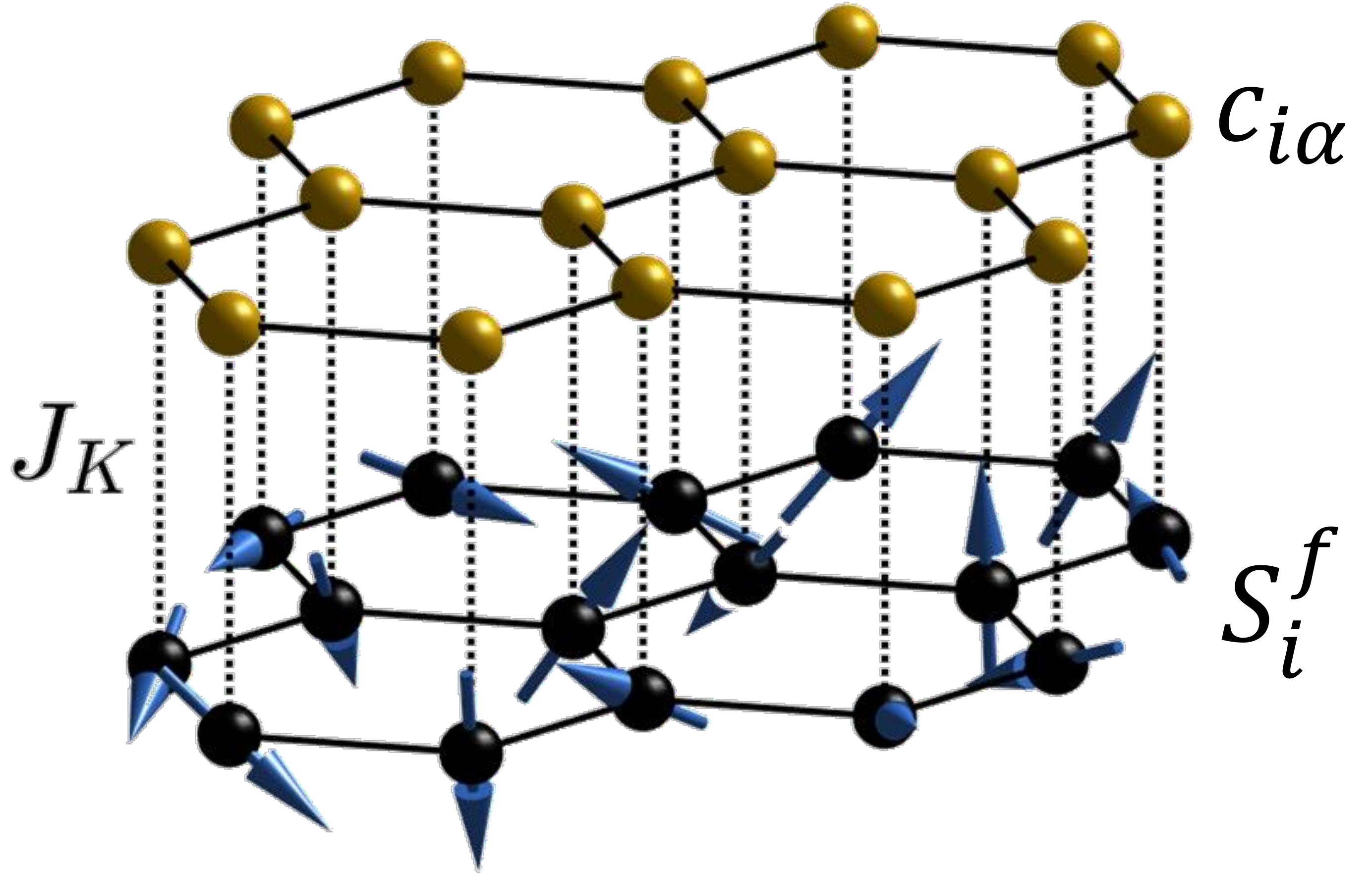}\label{fig:honeycomblattice}} ~~
	\subfloat[]{\raisebox{0.5ex}{\includegraphics[width=0.24\textwidth]{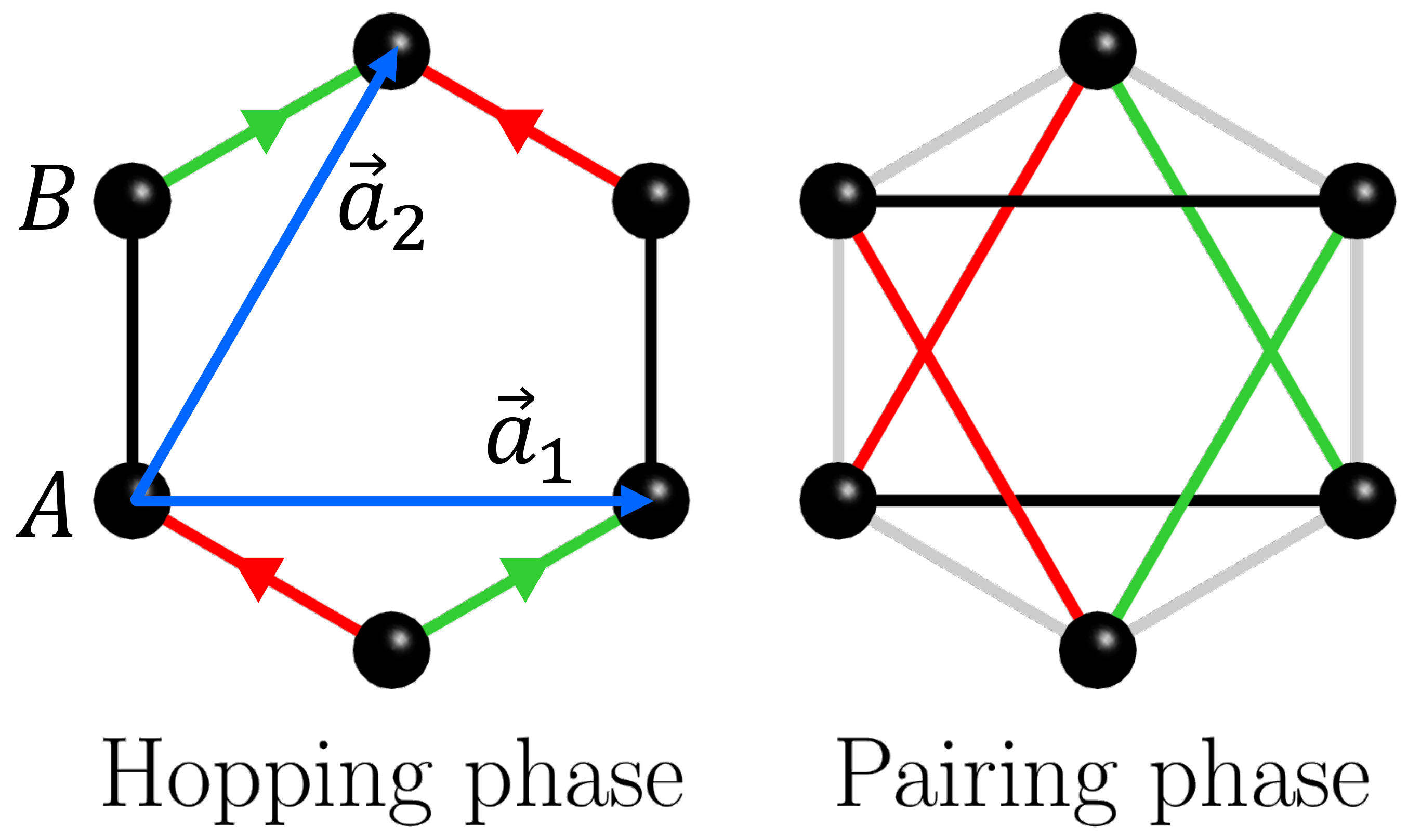}\hspace*{1em}\label{fig:phase_real}}} \\
	\captionsetup{oneside,margin={0.28cm,0cm}}
	\subfloat[]{\raisebox{-10ex}{\includegraphics[width=0.48\textwidth]{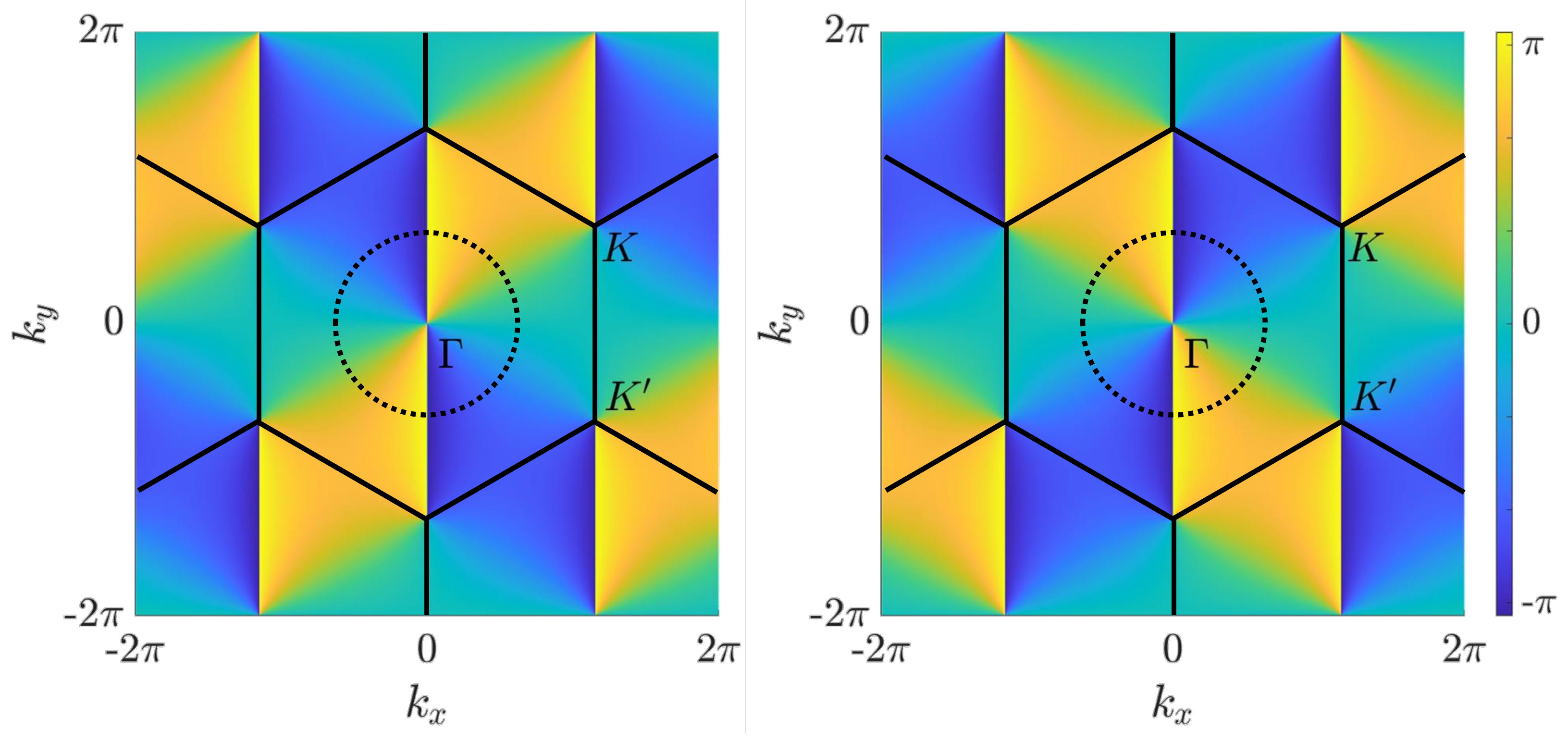}}\label{fig:phase_momentum}}
	\caption{(a) Kondo-Heisenberg model on the honeycomb lattice. Upper and lower layers represent conduction electrons ($c_{i\alpha}$) and localized moments ($S_{i}^{f}$) layer respectively. Those layers are coupled by on-site Kondo coupling $J_{K}$ depicted by dotted lines. (b) Nearest-neighbor hopping (left) and next-nearest-neighbor pairing (right) phases of $d \pm id$ spin liquid. Black, red, and green color indicate $1$, $e^{2\pi i/3}$, and $e^{4\pi i/3}$ respectively, and arrows on the left figure show the corresponding electron hopping direction. $\vec{a}_{1}$ and $\vec{a}_{2}$ are primitive lattice vectors, and $A,B$ are two sublattices in the unit cell. (c) Next-nearest-neighbor pairing phase of $d \pm id$ spin liquid for $A$-sublattice (left) and $B$-sublattice (right) in momentum space. Dotted line in the hexagonal plaquette shows the Fermi surface of quarter-filled conduction electron.}
\end{figure}

In this section, we study the $J_{1}$-$J_{2}$ Kondo-Heisenberg model on the quarter-filled honeycomb lattice. The system consists of two honeycomb layers (see Fig. \ref{fig:honeycomblattice}). The upper layer is occupied by one itinerant electron per unit cell on average governed by nearest-neighbor tight binding Hamiltonian. On the other hand, the lower layer is occupied by one localized spin moment $\bm{S}^{f}$ with $|\bm{S}^{f}| = 1/2$ per site governed by the nearest-neighbor and the next-nearest-neighbor anti-ferromagnetic Heisenberg exchanges $J_{1},J_{2} > 0$. Those two layers are coupled by the on-site Kondo interaction $J_{K}$ whose Hamiltonian is given by
\begin{align}
\label{eq:J1J2KondoHeisenberg}
    \mathcal{H} 
    &= t\sum_{\braket{i,j},\alpha}\left(c_{i\alpha}^{\dagger}c_{j\alpha} + h.c.\right) -\mu\sum_{i,\alpha}c_{i\alpha}^{\dagger}c_{i\alpha} \nn \\
    &+J_{1}\sum_{\braket{i,j}}\bm{S}_{i}^{f}\cdot\bm{S}_{j}^{f} +J_{2}\sum_{\braket{\braket{i,j}}}\bm{S}_{i}^{f}\cdot\bm{S}_{j}^{f} \nn \\
    &+J_{K}\sum_{i,\alpha\beta}\left(\frac{1}{2}c_{i\alpha}^{\dagger}\bm{\sigma}_{\alpha\beta}c_{i\beta}\right)\cdot\bm{S}_{i}^{f},
\end{align}
where $c_{i\alpha}^{\dagger}$ ($c_{i\alpha}$) is the conduction electron creation (annihilation) operator at site $i$ on the upper layer with spin $\alpha,\beta = \uparrow,\downarrow$, $\mu$ is the chemical potential for conduction electrons, and $\bm{\sigma}_{\alpha\beta}$ are Pauli matrices. From now on, the conduction electron hopping parameter $t$ is set to be unity and all energy and temperature scales are measured in unit of $t = 1$.

In the decoupled limit $J_{K} = 0$, the upper layer is quarter-filled metallic phase having finite Fermi surface (dotted line in the Fig. \ref{fig:phase_momentum}). The lower layer is $J_{1}$-$J_{2}$ Heisenberg model on the honeycomb lattice whose ground state for frustrated regime with both $J_{1}$ and $J_{2}$, is still on debate: spiral\cite{PhysRevB.84.094424}, plaquette valence bond solid\cite{PhysRevB.88.165138}, magnetically disordered\cite{PhysRevB.85.060402}, and spin liquid\cite{PhysRevB.96.104401,PhysRevLett.107.087204}. Among those candidates, $d \pm id$ spin liquid ansatz, the ground state confirmed by variational Monte Carlo simulation, is of our interest to study the FL*. To proceed, we introduce a fermionic spinon operator $f_{i\alpha}$ which constitutes the localized spin operator $\bm{S}_{i}^{f} = f_{i\alpha}^{\dagger}\bm{\sigma}_{\alpha\beta}f_{i\beta}/2$. The spinon is charge-neutral but carries fractionalized quantum number spin-$1/2$. By employing the mean-field order parameters,
\begin{align}
\label{eq:OrderParameters}
    b_{i} &= \braket{f_{i\alpha}^{\dagger}c_{i\alpha}}, & 
    \rho_{i} &= \varepsilon_{\alpha\beta}\braket{c_{i\alpha}^{\dagger}f_{i\beta}^{\dagger}}, \nn \\
    \chi_{ij} &= \braket{f_{i\alpha}^{\dagger}f_{j\alpha}}, & 
    \eta_{ij} &= \varepsilon_{\alpha\beta}\braket{f_{i\alpha}^{\dagger}f_{i\beta}^{\dagger}},
\end{align}
the on-site Kondo Hamiltonian and localized spin exchange terms in Eq. \eqref{eq:J1J2KondoHeisenberg} for $d \pm id$ spin liquid becomes
\begin{align}
    \mathcal{H}_{K} &= J_{K}\sum_{i}\left(b_{i}f_{i}^{\dagger}c_{i} + \rho_{i}f_{i}c_{i} + h.c.\right) \label{eq:HK} \\
    \mathcal{H}_{J_{1}} &= J_{1}\sum_{\braket{i,j}\alpha}\left(\chi_{ij}f_{i\alpha}^{\dagger}f_{j\alpha} + h.c.\right) \label{eq:HJ1} \\
    \mathcal{H}_{J_{2}} &= J_{2}\sum_{\braket{\braket{i,j}}}\left(\eta_{ij}f_{i\uparrow}f_{j\downarrow} + h.c.\right) \label{eq:HJ2}.
\end{align}
In Eq. \eqref{eq:OrderParameters}, $\varepsilon_{\alpha\beta}$ is Levi-Civita symbol, and the repeated indices are summed. The Kondo hybridization order parameters $b_{i}$ and $\rho_{i}$ are restricted to the on-site, while the spinon hopping $\chi_{ij} = \chi e^{i\phi_{ij}}$ and pairing $\eta_{ij} =  \eta e^{i\theta_{ij}}$ run over nearest-neighbor and next-nearest-neighbor sites respectively. Here, $\chi$ and $\eta$ are self-consistently calculated real number, and $\phi_{ij}$ and $\theta_{ij}$ are given in Fig. \ref{fig:phase_real}. Note that the Lagrange multiplier for spinon always vanishes $\lambda_{i} = 0$ to enforce the condition $\braket{f_{i}^{\dagger}f_{i}} = 1$ on average. During the mean-field calculation, we keep $\phi_{ij}$ and $\theta_{ij}$ for whole range of $J_{1}$, $J_{2}$, and $J_{K}$ to investigate the phase transitions out of FL*. In the next section, we will explore the translation invariant saddle point solutions of Eq. \eqref{eq:OrderParameters} to Eq. \eqref{eq:HJ2}.

\section{Mean-field phase diagram}

\begin{figure}
    \centering
    \includegraphics[width = \linewidth]{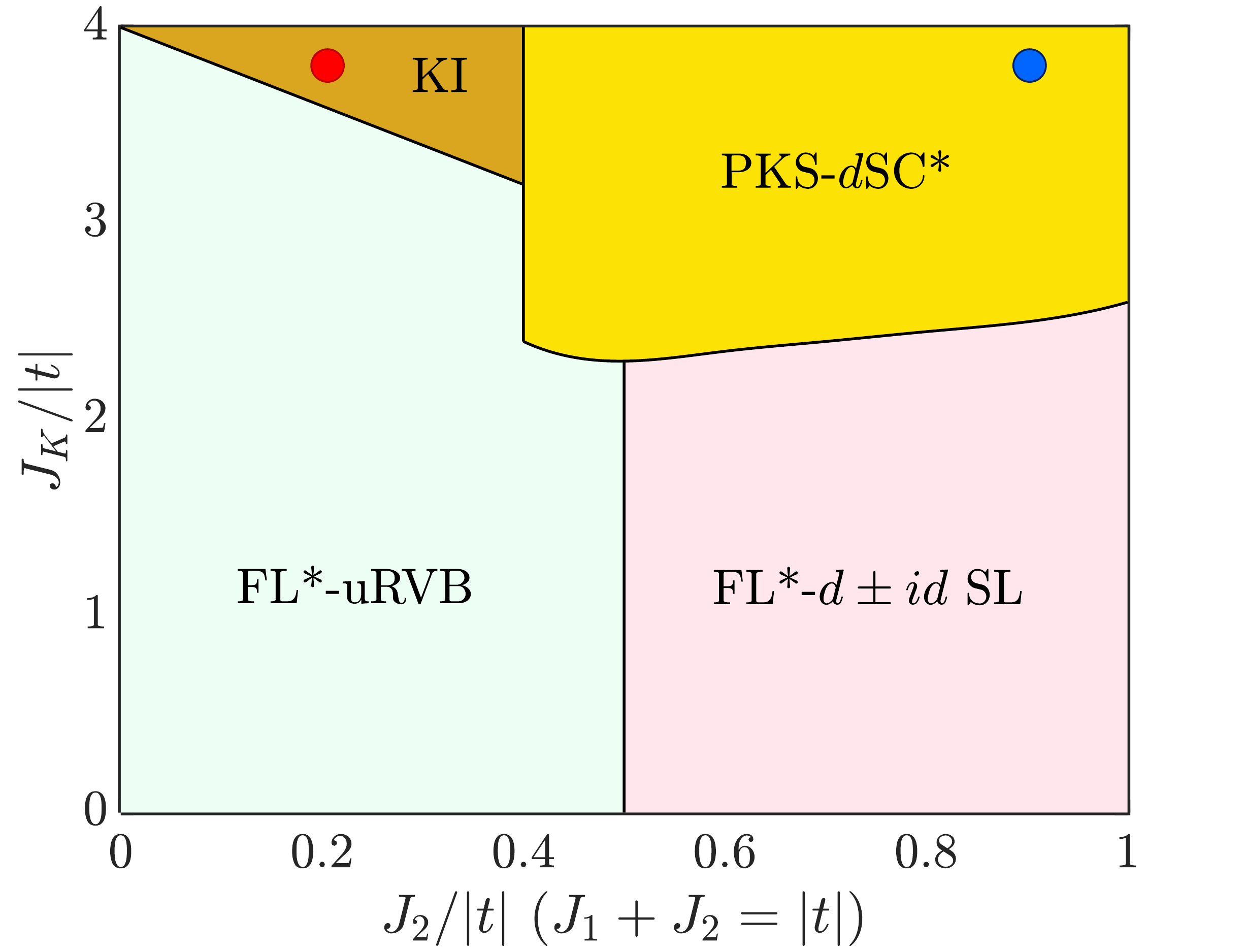}
    \caption{The mean-field phase diagram of $J_{1}$-$J_{2}$ Kondo-Heisenberg model on the quarter-filled honeycomb lattice. Each abbreviation in the phase diagram indicates fractionalized Fermi liquid - uniform resonant valence bond (FL*-uRVB), fractionalized Fermi liquid - $d \pm id$ spin liquid (FL*-$d\pm id$ SL), partial Kondo screening - chiral $d$-wave fractionalized superconductor (PKS-$d$SC*), and Kondo insulator (KI). Blue and red dots which are the colors for thermal conductance plot and add markers in Fig. \ref{fig:thermal_conductance} as well are the parameters to plot Fig. \ref{fig:thermal_conductance}.}
    \label{fig:PhaseDiagram}
\end{figure}

\begin{figure}
	\subfloat[fractionalized Fermi liquid - uniform resonant valence bond]{\includegraphics[width=0.48\textwidth]{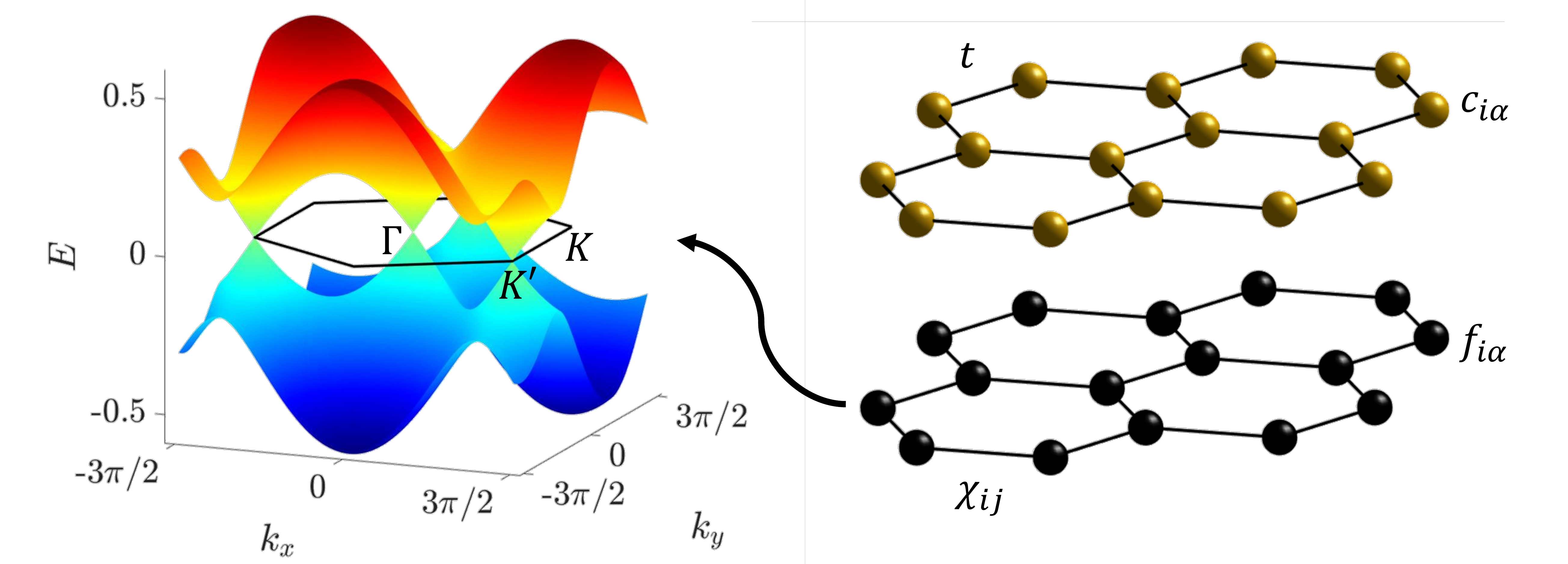}\label{fig:uRVB}} \\
	\subfloat[fractionalized Fermi liquid - $d \pm id$ spin liquid]{\includegraphics[width=0.48\textwidth]{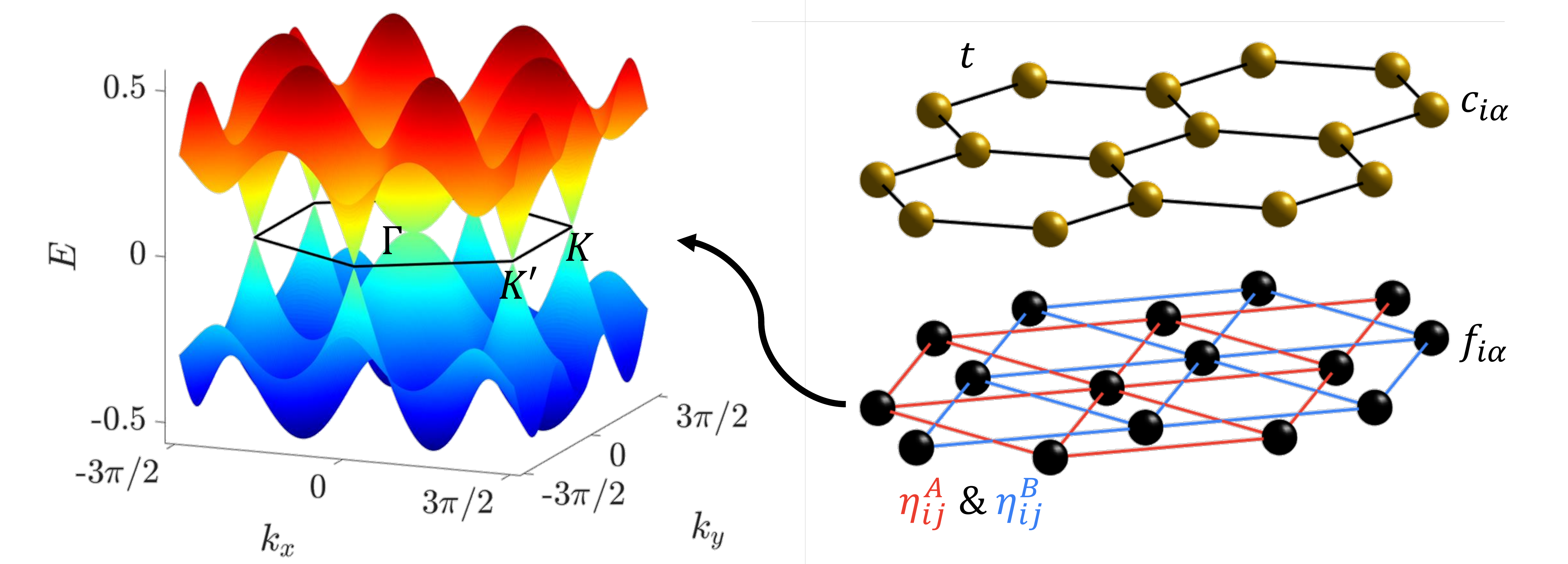}\label{fig:did}} \\
	\subfloat[Partial Kondo screening - chiral $d$-wave fractionalized superconductor]{\includegraphics[width=0.48\textwidth]{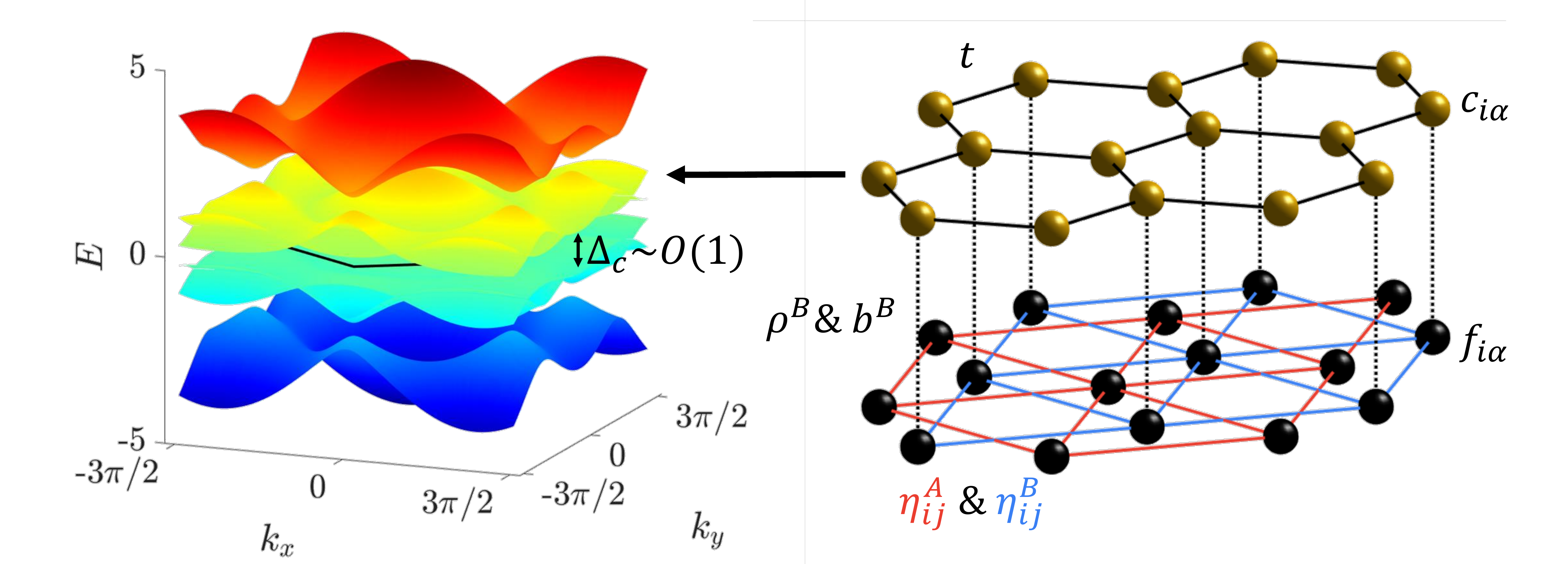}\label{fig:PKS}}
	\caption{The spinon band structure of (a) Fractionalized Fermi liquid - uniform resonant valence bond ($J_{1} = 1, J_{2} = 0$) and (b) Fractionalized Fermi liquid - $d \pm id$ spin liquid ($J_{1} = 0, J_{2} = 1$). The band structure of $d \pm id$ spin liquid is doubly degenerate. (c) The electron band structure of partial Kondo screening - chiral $d$-wave fractionalized superconductor ($J_{1} = 0, J_{2} = 1, J_{K} = 4$). The $A$-sublattice spinon spectrum is the same as Fig. \ref{fig:did}. $\Delta_{c} \sim O(1)$ is the superconducting gap induced from Kondo hybridization with $B$-sublattice of spinon layer. The detail analysis about $\Delta_{c}$ is discussed in section \ref{sec:analysis}. The black lines in the left column represent first Brillouin zone boundary.}
\end{figure}

Below the critical Kondo value $0 \leq J_{K} < \left.J_{K}\right|_{c}$, no electron-spinon hybridization occurs $b_{i} = \rho_{i} = 0$, which preserves the emergent gauge symmetry mediating the spinons. In this case, the mean-field solution favors different FL* phases depending on the ratio between $J_{1}$ and $J_{2}$. For $J_{2}/J_{1} \lesssim 1$ where the frustration effect is less dominant, the spinon propagates on the uniformly finite spinon hopping order parameter strength $|\chi_{ij}| = \chi$ while the next-nearest-neighbor spinon pairing vanishes $\eta_{ij} = 0$, called fractionalized Fermi liquid - uniform resonant valence bond phase (FL*-uRVB). The schematic order parameter configuration and corresponding spinon band structure are shown in the Fig. \ref{fig:uRVB}. In the band structure, there are two Dirac cones at the high symmetry points: $\Gamma$ and $K'$. Those are protected by the product of inversion $\mathcal{P}$ and time-reversal $\mathcal{T}$ symmetries. Although the $\mathcal{PT}$-symmetric mean-field ansatz does not preserve $\mathcal{P}$ and $\mathcal{T}$ symmetry separately, the Gutzwiller projected wave function is symmetric under both $\mathcal{P}$ and $\mathcal{T}$ symmetries\cite{PhysRevB.96.104401}.

Increasing $J_{2}/J_{1}$, the first order phase transition occurs toward fractionalized Fermi liquid - $d \pm id$ spin liquid phase (FL*-$d \pm id$ SL) (see Fig. \ref{fig:did}). In this case, the nearest-neighbor spinon hopping order parameter vanishes $\chi_{ij} = 0$, but has finite next-nearest-neighbor spinon pairing order parameters $\eta_{ij} \neq 0$. As shown in the right side of Fig. \ref{fig:did}, two sublattices $A$ and $B$ on the spinon layer are completely decoupled so that the spinon spectrum in the Fig. \ref{fig:did} is doubly degenerate. Even though the spinon band structure of two sublattices are completely equivalent, The chiral structure around the gapless points are opposite. In the Fig. \ref{fig:phase_momentum}, we plot the phase of next-nearest-neighbor pairing term for each sub-lattice on the momentum space $\eta_{A}(\bm{k}) = |\eta_{A}{\bm{k}}|\exp i\theta_{A}(\bm{k})$ and $\eta_{B}(\bm{k}) = |\eta_{B}{\bm{k}}|\exp i\theta_{B}(\bm{k})$. At the $\Gamma$-point, quadratic band touching occurs and the winding number around the $\Gamma$-point is $w_{\Gamma} = \pm 2$, or equivalently $d \pm id$ chiral structure, on the each sublattice. In addition, there exist two Dirac cones at $K$ and $K'$ points with winding number $w_{K,K'} = \mp 1$, or equivalently $p \mp ip$ chiral structure since the winding number over whole Brillouin zone should be 0. Those gapless points with finite chiral structure are protected by $\mathcal{PT}$-symmetry similar to FL*-uRVB. 

For sufficiently large $J_{K}$ with $J_{2}/J_{1} < 2/3$, the spinons and conduction electrons evenly take part in the hybridization ($b_{i} \neq 0$ and $\rho_{i} \neq 0$ for all $i$) to become a Kondo insulator (KI). As the electron-spinon hybridization order parameters $b_{i}$ and $\rho_{i}$ increase, the $\chi_{ij}$ and $\eta_{ij}$ between hybridized sites naturally decrease since the local spin fluctuation is suppressed rather than generating the spinon hopping or pairing.

However, if considerable frustrated Heisenberg exchange $J_{1} \lesssim J_{2}$ exists, the one-sublattice, say $B$-sublattice, of the honeycomb sites are spontaneously covered by the spinon-electron singlets, and the other sublattice is not by virtue of the commensurate filling of conduction electrons (see Fig. \ref{fig:PKS}). Therefore, $b_{B},\rho_{B} \neq 0$ and the local moments on $A$-sublattice are decoupled from the Kondo singlet sites ($b_{A} = \rho_{A} = \chi_{ij} = 0$) with its own $d$-wave spin liquid solution optimizing Eq. \eqref{eq:HJ2} ($\eta_{A} \neq 0$). This quantum phenomena is called partial Kondo screening. Although the spinon pairing is always finite on $A$-sublattice, the magnitude of $B$-sublattice pairing diminishes as the hybridization is solidified. It obviously vanishes for infinite $J_{K}$ limit, but we find $\eta_{B}, b_{B}, \rho_{B} \neq 0$ solution for finite Kondo coupling strength distinct from the Kondo limit. In this case, the Cooper pairing is induced on the hybridized sites whose gap structure is inherited from $d - id$ pairing on $B$-sublattice with preserving chirality. Since $A$-sublattice still keeps its own $d$-wave spin liquid induced by partial Kondo screening, the deconfined spinons coexists with the chiral $d$-wave superconductor, or partial Kondo screening - chiral $d$-wave fractionalized superconductor (PKS-$d$SC*).

\section{Thermal conductance behaviours}

\begin{figure}
    \centering
    \includegraphics[width = \linewidth]{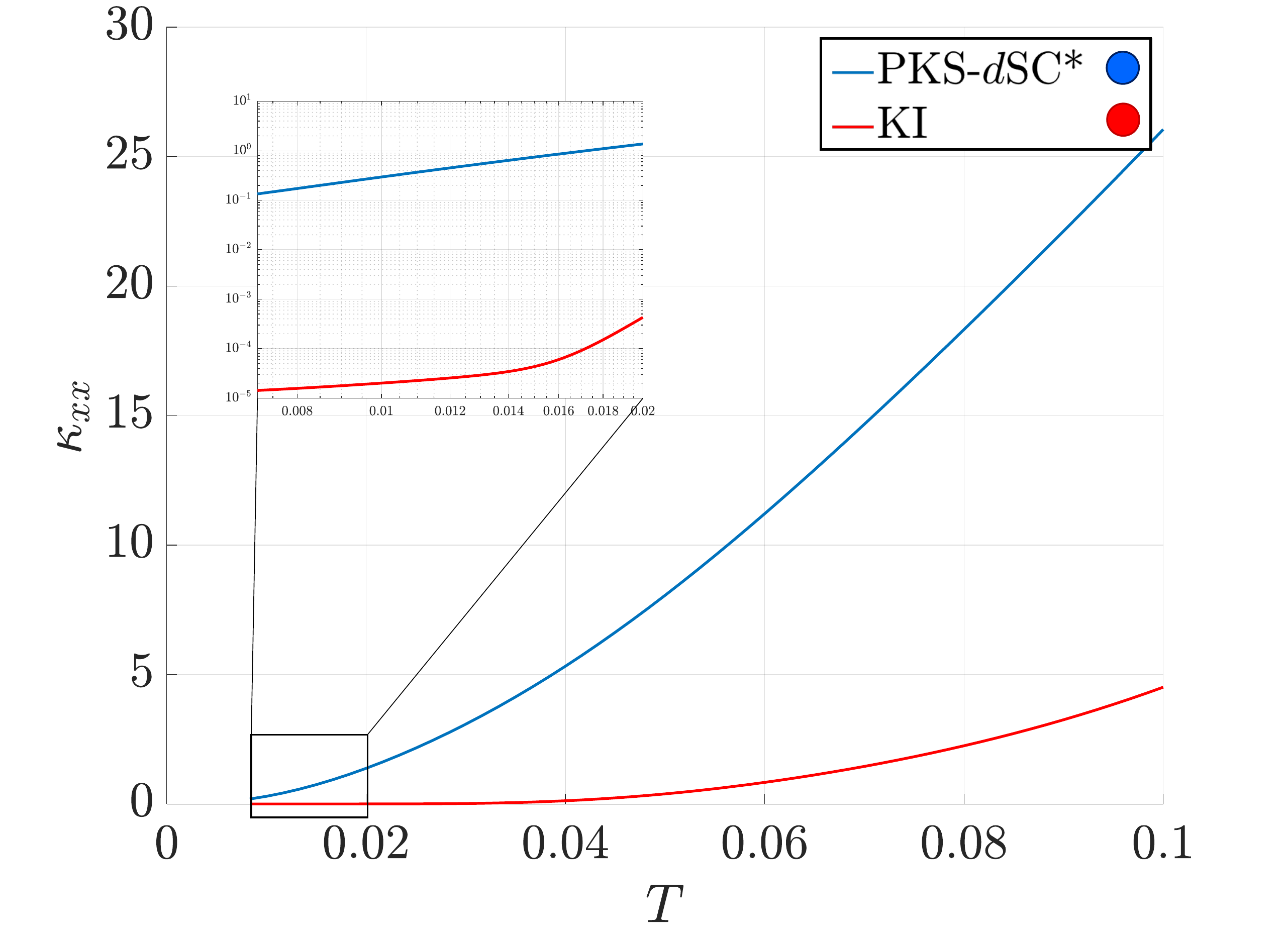}
    \caption{Longitudinal thermal conductance - temperature plot for partial Kondo screening - chiral $d$-wave fractionalized superconductor (PKS-$d$SC*) and Kondo insulator (KI). The inset shows log-log plot for low temperature limit. The linear relation between $\log \kappa_{xx}$ and $\log T$ indicates that the low temperature thermal conductance for PKS-$d$SC* has power law behaviour. The parameters we choose to plot are dotted blue (PKS-$d$SC*) and red (KI) in Fig. \ref{fig:PhaseDiagram}.}
    \label{fig:thermal_conductance}
\end{figure}

The existence of the gapless excitation and its dispersion specifies the thermodynamics behaviours. The FL*-uRVB and FL*-$d \pm id$ SL contains both the electron and gapless spinon excitation. However, such excitation is frozen in the KI. The PKS-$d$SC* only contains gapless spinon excitation and the electron sector is gapped out by electron-spinon hybridization partially inducing superconductivity. This affects the heat transport, especially the thermal conductance which is of our interest. The thermal conductance tensor $\kappa_{\mu\nu}$ ($\mu,\nu = x,y$ in $d = 2$ spatial dimension) at temperature $T$ is given by,
\begin{align}
    \kappa_{\mu\nu}(T) &= \frac{1}{T^{2}}\left(L_{\mu\nu}^{22}(T) - \frac{[L_{\mu\nu}^{12}(T)]^{2}}{L_{\mu\nu}^{11}(T)}\right),\label{eq:ThermalConductance}
\end{align}
where the $L_{\mu\nu}^{11}$, $L_{\mu\nu}^{12}$, and $L_{\mu\nu}^{22}$ are defined as,
\begin{align}
    L_{\mu\nu}^{11}(T) &= \int_{-\infty}^{\infty}dE[-\partial_{E}f(E,T)]\zeta_{\mu\nu}(E,T) \nn \\
    L_{\mu\nu}^{12}(T) &= \int_{-\infty}^{\infty}dEE[-\partial_{E}f(E,T)]\zeta_{\mu\nu}(E,T) \nn \\
    L_{\mu\nu}^{22}(T) &= \int_{-\infty}^{\infty}dEE^{2}[-\partial_{E}f(E,T)]\zeta_{\mu\nu}(E,T).
\end{align}
Here, $f(E,T) = 1/(e^{\beta E} + 1)$ ($\beta = 1/T$ with $k_{B} = 1$) is Fermi-Dirac distribution function and $\zeta_{\mu\nu}(E,T)$ is,
\begin{align}
    \zeta_{\mu\nu}(E,T) &= \frac{T}{\pi N_{b}N_{c}}\sum_{b=1}^{N_{b}}\sum_{\bm{k}}v_{\mu}^{(b)}(\bm{k})v_{\nu}^{(b)}(\bm{k})[\Im G^{(b)}(\bm{k},E)]^{2}.
\end{align}
Here, $G^{(b)}(\bm{k},E)$ is a Green's function for $b$-th band, $v_{\mu}(\bm{k})$ is $\mu$-th component of group velocity, $N_{b}$ is the number of band, and $N_{c}$ is the number of grid we set in the numerical calculation.

In Fig. \ref{fig:thermal_conductance}, the low temperature behaviour of longitudinal thermal conductance $\kappa_{xx}$ is plotted. Blue and red line represent the thermal conductance - temperature relation of PKS-$d$SC* and KI, that are marked as blue and red dots in Fig. \ref{fig:thermal_conductance} respectively . The inset shows log-log plot of low temperature ($0.005 < T < 0.02$). In KI phase, it exhibits the exponential suppression due to the finite gap induced by Kondo hybridization. Meanwhile, PKS-$d$SC* features almost linear relations between logarithm of thermal conductance and logarithm of temperature, which implies the power law of thermal conductance, even though it also has a finite gap in conduction electron layers induced by Kondo hybridization. This anomalous thermal conductance behaviour comes from the gapless excitation of spinons which do not couple with electron layer. Note that the thermal conductance behaviours of conventional or gapped chiral superconductor is exponential decaying due to its gap spectrum.

\section{Analysis}
\label{sec:analysis}
In the mean-field calculation, FL*-uRVB and FL*-$d \pm id$ SL are shown to be stable up to critical Kondo coupling strength (see Fig. \ref{fig:PhaseDiagram}). In this regime, the local spin moments strongly fluctuate but is decoupled from the itinerant electrons. The robustness of FL* compared to the conventional ordered magnet is generically guaranteed by the topological order. Here, we discuss the influence of Kondo interactions to our mean-field solutions at $J_{k} = 0$. In the continuum $(d + 1)$-dimension, the generic onsite coupling which respects the global $U(1)$-symmetry (local $Z_{2}$-symmetry) of $c$- ($f$-) fermion is
\begin{align}
    \mathcal{S}_{\text{int}} &= \sum_{\alpha\beta\gamma\delta}J_{\alpha\beta\gamma\delta}\int d^{d}xd\tau\left(c_{\alpha}^{\dagger}(\bm{x})c_{\beta}(\bm{x})\right)\left(f_{\gamma}^{\dagger}(\bm{x})f_{\delta}(\bm{x})\right). \label{eq:SK}
\end{align}
Let's assume that the dispersion of electron and spinon are linearized, $\varepsilon_{\bm{k}}^{c,f} \sim v^{c,f}|\bm{k}|$ on the momentum space. In the strong coupling limit $J_{\alpha\beta\gamma\delta} \to \infty$, it obviously condensate $\braket{c_{\alpha}^{\dagger}(\bm{x})f_{\beta}(\bm{x})} \neq 0$ and $\braket{c_{\alpha}^{\dagger}(\bm{x})f_{\beta}^{\dagger}(\bm{x})} \neq 0$ which break the global $U(1)$- and $Z_{2}$-symmetries. In our context, this correspond to Kondo insulator. To examine whether the arbitrary small interaction $J_{\alpha\beta\gamma\delta}$ leads to this picture or not, we employ the dimensional analysis on Eq. \eqref{eq:SK}. In the linearized action, the scaling dimensions of $c$ and $f$ fermions are $[c] = [f] = d/2$. As a result, the scaling dimension of $\mathcal{S}_{\text{int}}$ is $[J_{\alpha\beta\gamma\delta}] = 1-d$, which implies the irrelevance of Eq. \eqref{eq:SK} for the spatial dimension larger than 1 at the tree level. If the local spin operators are fractionalized into deconfined spinons, the power counting is applicable to the Kondo interaction in Eq. \eqref{eq:H}. This argument supports the reliability of our mean-field results in Fig. \ref{fig:PhaseDiagram} and clarifies the Kondo-breakdown for small $J_{K}$ separated from the heavy Fermi liquid.

Now, we discuss the chiral gap structure on the itinerant electrons induced by partial Kondo screening. With the itinerant electron side, the mean-field decoupled Hamiltonian Eq. \eqref{eq:HK} to Eq. \eqref{eq:HJ2} can be arranged as,
\begin{gather}
    \mathcal{H} = \sum_{\bm{k}}
    \begin{pmatrix}
    \Psi_{c}(\bm{k}) \\
    \Psi_{f}(\bm{k})
    \end{pmatrix}^{\dagger}
    \begin{pmatrix}
    H_{c}(\bm{k}) & K \\
    K^{\dagger} & H_{f}(\bm{k})
    \end{pmatrix}
    \begin{pmatrix}
    \Psi_{c}(\bm{k}) \\
    \Psi_{f}(\bm{k})
    \end{pmatrix}
    \label{eq:H}, \\
    \Psi_{c,f}(\bm{k}) = 
    \begin{pmatrix}
    c,f_{A\uparrow}(\bm{k}) \\
    c,f_{A\downarrow}^{\dagger}(-\bm{k}) \\
    c,f_{B\uparrow}(\bm{k}) \\
    c,f_{B\downarrow}^{\dagger}(-\bm{k})
    \end{pmatrix} \equiv 
    \begin{pmatrix}
    \psi_{c,f}^{\text{U}}(\bm{k}) \\
    \psi_{c,f}^{\text{S}}(\bm{k})
    \end{pmatrix}
\end{gather}
where $H_{c}(\bm{k})$ and $H_{f}(\bm{k})$ are electron and spinon Hamiltonian matrix at $\bm{k}$ respectively. $H_{c}(\bm{k})$ and $H_{f}(\bm{k})$ are connected via translation invariant onsite Kondo coupling matrix $K$, thus momentum independent. Beyond the critical Kondo coupling strength with frustrated regime, we know that one sublattice spontaneously hybridizes with electron layers. Let's assume that the hybridized sublattice is $B$-sublattice (see Fig. \ref{fig:PKS}). Then we can divide $A$- and $B$-sublattice as Kondo unscreened and screened sites, denoted as U and S in the superscript. From now on, we omit the momentum dependency of Hamiltonian matrices and basis operators unless there is any confusion. 

When $J_{K} \geq J_{K}^{c}$, the Kondo coupling matrix $K$ and spinon Hamiltonian $H_{f}$ becomes,
\begin{align}
    \Psi_{c}^{\dagger}K\Psi_{f} &= 
    \begin{pmatrix}
    \psi_{c}^{\text{U}} \\
    \psi_{c}^{\text{S}}
    \end{pmatrix}^{\dagger}
    \begin{pmatrix}
    \bm{K}^{\text{U}} = \bm{0} & \bm{0} \\
    \bm{0} & \bm{K}^{\text{S}}
    \end{pmatrix}
    \begin{pmatrix}
    \psi_{f}^{\text{U}} \\
    \psi_{f}^{\text{S}}
    \end{pmatrix}\label{eq:K} \\
    \Psi_{f}^{\dagger}H_{f}\Psi_{f} &= 
    \begin{pmatrix}
    \psi_{f}^{\text{U}} \\
    \psi_{f}^{\text{S}}
    \end{pmatrix}^{\dagger}
    \begin{pmatrix}
    \bm{H}_{f}^{\text{U}} & \bm{h}_{f} = \bm{0} \\
    \bm{h}_{f}^{\dagger} = \bm{0} & \bm{H}_{f}^{\text{S}}
    \end{pmatrix}
    \begin{pmatrix}
    \psi_{f}^{\text{U}} \\
    \psi_{f}^{\text{S}}
    \end{pmatrix}\label{eq:Hf},
\end{align}
where $\bm{K}^{\text{S}}$ is non-zero matrix which hybridizes electron and spinon on the Kondo screened sites. We note that the Kondo matrix for unscreened sites is $\bm{K}^{\text{U}} = \bm{0}$ as the name unscreened depicted. As we discussed earlier, finite order parameters for Kondo coupling on the Kondo screened sites implies that the spinon on the Kondo screened sites prefer interacting with electrons rather than spinons. Thus, it makes $\bm{h}_{f} = \bm{0}$ making spinons on the Kondo unscreened sites completely decoupled to electron layer and spinons on the Kondo screened sites.

Let us first calculate the induced electron Hamiltonian driven by PKS. We integrate out the Eq. \eqref{eq:Hf} from Eq. \eqref{eq:H} and obtain the effective electron action $\mathcal{S}_{c}^{\text{eff}}$ given as,
\begin{align}
    \mathcal{S}_{c}^{\text{eff}} &= \sum_{\bm{k},i\omega_{n}}\bar{\Psi}_{c}\left(G_{c}^{-1} - KG_{f}K^{\dagger}\right)\Psi_{c}
    \label{eq:EffectiveElectronAction},
\end{align}
where $G_{c,f} = \left(-i\omega_{n} + H_{c,f}\right)^{-1}$ is Green's function of electron and spinon, and $\omega_{n} = (2n+1)\pi/\beta$ is fermionic Matsubara frequency. At low temperature, the induced term of Eq. \eqref{eq:EffectiveElectronAction} can be outlined by substituting $\omega_{n} = 0$ as,
\begin{align}
\label{eq:lowtemKGK}
    \Psi_{c}^{\dagger}K\tilde{G}_{f}K^{\dagger}\Psi_{c} = 
    \begin{pmatrix}
    \psi_{c}^{\text{U}} \\
    \psi_{c}^{\text{S}}
    \end{pmatrix}^{\dagger}
    \begin{pmatrix}
    \bm{0} & \bm{0} \\
    \bm{0} & \bm{K}^{\text{S}}[\bm{H}_{f}^{\text{S}}]^{-1}\bm{K}^{\text{S}\dagger}
    \end{pmatrix}
    \begin{pmatrix}
    \psi_{c}^{\text{S}} \\
    \psi_{c}^{\text{U}}
    \end{pmatrix}.
\end{align}
Let's assume that the Kondo screened sites form a translation invariant lattice structure with a single sublattice, such as triangular or square lattice. Then $\bm{H}_{f}^{\text{S}}$ and $\bm{K}^{\text{S}}$ can be written as $2 \times 2$ matrix form in the momentum space as,
\begin{align}
\label{eq:KSHS}
    \psi_{c}^{\text{S}\dagger}\bm{K}^{\text{S}}\psi_{f}^{\text{S}} &= 
    \begin{pmatrix}
    c_{\bm{k}\uparrow}^{\dagger} & c_{-\bm{k}\downarrow}
    \end{pmatrix}
    \begin{pmatrix}
    b & \rho^{*} \\
    \rho & -b^{*}
    \end{pmatrix}
    \begin{pmatrix}
    f_{\bm{k}\uparrow} \\
    f_{-\bm{k}\downarrow}^{\dagger}
    \end{pmatrix} \\
    \psi_{f}^{\text{S}\dagger}\bm{H}_{f}^{\text{S}}\psi_{f}^{\text{S}} &=
    \begin{pmatrix}
    f_{\bm{k}\uparrow}^{\dagger} & f_{-\bm{k}\downarrow}
    \end{pmatrix}
    \begin{pmatrix}
    \xi_{f} & \eta^{*} \\
    \eta & -\xi_{f}
    \end{pmatrix}
    \begin{pmatrix}
    f_{\bm{k}\uparrow} \\ 
    f_{-\bm{k}\downarrow}^{\dagger}
    \end{pmatrix} \nn,
\end{align}
where $b$, $\rho$, and $\eta$ are defined in Eq. \eqref{eq:OrderParameters}. Inserting Eq. \eqref{eq:KSHS} into Eq. \eqref{eq:lowtemKGK}, the induced electron Hamiltonian on the Kondo screened sites near the Fermi surface becomes
\begin{widetext}
\begin{align}
    \bm{H}_{c}^{\text{S}} &\equiv 
    -\bm{K}^{\text{S}}[\bm{H}_{f}^{\text{S}}]^{-1}\bm{K}^{\text{S}\dagger} 
    \equiv 
    \begin{pmatrix}
    \xi_{c}^{\text{S}} & \Delta_{c}^{\text{S}*} \\
    \Delta_{c}^{\text{S}} & -\xi_{c}^{\text{S}}
    \end{pmatrix} = \frac{1}{\xi_{f}^{2}+|\eta|^{2}}
    \begin{pmatrix}
    -2\Re[\eta^{*} b \rho]-\xi_{f}(|b|^{2} - |\rho|^{2}) & b^{2}\eta^{*} - \rho^{*2}\eta - 2b\rho^{*}\xi_{f} \\
    b^{*2}\eta - \rho^{2}\eta^{*} - 2b^{*}\rho\xi_{f} & 2\Re[\eta^{*} b\rho]+\xi_{f}(|b|^{2} - |\rho|^{2})
    \end{pmatrix} 
    \label{eq:KGK},
\end{align}
\end{widetext}
where $\Re$ is the real part, and $\xi_{c}^{\text{S}}$ and $\Delta_{c}^{\text{S}}$ are induced electron hopping and pairing term on the Kondo screened sites.

Close to the Fermi surface where the denominator is non-vanishing, we focus on the induced Cooper pairing in the itinerant electron layer. In our saddle point solutions, it turns out that $|\rho| < |b|$ thus the off-diagonal $\Delta_{c}^{\text{S}}$ is largely contributed from $b^{*2}\eta$. Therefore, the chirality of induced electron pairing has the same chiral structure of spinons on the hybridized sublattice. Furthermore, the magnitudes of order parameters in our numerical solution are given as $\xi_{f} = 0$, $|b| \sim O(10^{-1})$, $|\eta| \sim O(10^{-2})$, and $|\rho| \sim O(10^{-3})$ and result in $|\xi_{c}^{\text{S}}| \sim O(10^{-2})$ and $|\Delta_{c}^{\text{S}}| \sim O(1)$ which implies that the superconducting gap has an order of 1. In the Fig. \ref{fig:PKS}, $\Delta_{c} \sim O(1)$ which is perfectly matches with our analytical calculation.

\section{Conclusions}
In this work, we study the $J_{1}$-$J_{2}$ Kondo-Heisenberg model at quarter-filled conduction electrons on the honeycomb lattice. Based on the slave-fermion mean-field approach, new fractionalized phase, called PKS-$d$SC* phase, out of weak-coupling FL* regime is proposed by turning on the Kondo coupling collaborated with frustrated RKKY-exchanges. This phase appears for a wide range of mean-field phase diagram having a fractionalized spinon excitation due to the frustration effect endowed with the unbalanced filling. We then discuss the stability of fractional excitation against the Kondo exchange, the chirality of the induced superconductivity, and the thermal conductance behaviours for each phase as the temperature varies.

For sufficiently large $J_{K} \gg J_{1},J_{2}$, we check that the partial Kondo screening occurs when $J_{2}/J_{1} \gtrsim 2$ by exact diagonalization up to $18$ sites honeycomb lattice with open-boundary condition. Therefore, it will be interesting to study the partial Kondo screening out of FL* beyond the mean-field regime on the different frustrated lattice systems. Furthermore, we can also consider the Lifshitz transition controlled by filling factor. As can be seen in Fig. \ref{fig:phase_momentum}, there is $p \mp ip$ pairing structure around the $K$ and $K'$ points for FL*-$d \pm id$ SL. Therefore, if the electron filling factor becomes larger than $1/3$ thus the Fermi surface surrounds $K$ and $K'$ points keeping partial Kondo screening, we expect the Lifshitz transition from PKS-$d$SC* to PKS-$p$SC*, which will be left for the future work.

\subsection*{Acknowledgement}
We thank Tarun Grover for useful discussions. This work is supported by National Research Foundation Grant (NRF-
2020R1F1A1073870, NRF-2020R1A4A3079707)).

\bibliography{sample}

\end{document}